# Re-sampling of inline holographic images for improved reconstruction resolution

S.G. Podorov[1,2,*], A.I. Bishop[2], D.M. Paganin[2] and K.M. Pavlov[2,3]

[1] *Institut für Röntgenphysik, Universität Göttingen, Friedrich-Hund-Platz 1, 37077 Göttingen, Germany*

[2] *School of Physics, Monash University, Wellington RD, Clayton, VIC 3800, Australia*

[3] *Physics and Electronics, School of Science and Technology, University of New England, NSW 2351, Australia*

*Corresponding author: Sergey.Podorov@phys.uni-goettingen.de*

**Abstract:** Digital holographic microscopy based on Gabor in-line holography is a well-known method to reconstruct both the amplitude and phase of small objects. To reconstruct the image of an object from its hologram, obtained under illumination by monochromatic scalar waves, numerical calculations of Fresnel integrals are required. To improve spatial resolution in the resulting reconstruction, we re-sample the holographic data before application of the reconstruction algorithm. This procedure amounts to inverting an interpolated Fresnel diffraction image to recover the object. The advantage of this method is demonstrated on experimental data, for the case of visible-light Gabor holography of a resolution grid and a gnat wing.

1. **Introduction**

Digital holographic microscopy is a very fast and promising optical tool for investigation of small objects [1]. The digital reconstruction of inline Gabor holograms may be based on the methods that use either one or two fast Fourier transform (FFT) operations [2]. For this numerical procedure the number of pixels in the detector plane and object plane is the same. When holograms are produced using plane wave illumination, the resolution of the method is limited by pixel size of the detector. As the pixel size (generally 1-20 μm) is very often larger than the theoretical limit given by the Abbe estimation, many authors work on numerical procedures to improve the resolution of the method or to obtain subpixel information for the studied object.

Some of these algorithms are based on minimization of the normalized root-mean-square error between reconstructed images and Fourier transforms of the images [3]. Other algorithms are based on several measurements done with small detector shifts at a subpixel scale [4]. In [5] a new sub-pixel detector for X-corners in camera calibration targets was proposed. This algorithm consists of a new X-corner operator, used together with a second-order Taylor series approximation describing the local intensity profile around the X-corner. The sub-pixel position of the X-corner can be determined directly by calculating the saddle point of this polynomial approximation.

In [6] a new class of reconstruction algorithms, known as "imaging-consistent" reconstructions, was introduced. They are fundamentally different from traditional approaches. Here, the authors [6] treat image values as area samples generated by non-overlapping integrators, which are consistent with the image formation process in CCD cameras. They obtained excellent results by formulating reconstruction as a two-stage process: image restoration, followed by the application of the point spread function (PSF) of the imaging sensor. Efficient local techniques for image restoration are derived to invert the effects of the PSF and to estimate the underlying image that passed through the sensor. Boult and Wolberg [6] define the imaging-consistency constraint which requires the approximate reconstruction to be the exact reconstruction for functions in the specified admissible class of functions. This class of functions is defined by bounding the maximum absolute value of the second derivative of the underlying continuous signal. The error of the algorithms [6] can be shown to be at most twice the error of the optimal algorithm for a wide range of optimality constraints.

The two-step digital image correlation method is developed with a more stable and credible calculating technology in [7]; this consists of the simple searching method and correlation iterative method. An interesting subpixel image reconstruction algorithm was described in [8] without references to holography. The paper [9] proposes an iterative error-energy reduction algorithm to reconstruct the high-resolution (alias-free) output image that utilizes a correlation method to estimate the subpixel shifts among a sequence of low-resolution aliased imagery. The proposed super-resolution image reconstruction algorithm provides a possibility to produce high-resolution images by using low-resolution images from existing low-cost imaging devices. In paper [10], Chen *et al.* have shown how to model a subpixel registration algorithm based on digital correlation (DC) so as to provide significant robustness to intensity variations and noise corruption. Based on the different statistical properties between signal and noise, these authors [10] have derived an analytical expression for the cross-correlation coefficient in the presence of noise, and have further registered the subpixel translation by introducing Taylor series expansions for each frame. In [11] Le and Seetharaman proposed to use a videocamera to register shifts of the object and apply this to register subpixel information of the images.

In this paper we propose an alternative approach to the above listed methods. Here we interpolate not the reconstructed image, but the inline Gabor holographic images before the application of reconstruction algorithm. The effectiveness of the method is shown using experimentally recorded images of grid and gnat wings. The method allows us to improve the spatial resolution of the reconstructed images by a factor of three to five times.

## 2. Theoretical part

Gabor in-line holography is a well-known method for reconstruction of images from their Fresnel diffraction patterns. The goal is to reconstruct the complex scalar wave-field (image) $E(x,y, z = 0)$ from the Gabor hologram corresponding to intensity distribution $I_G(x,y, z = L)$, recorded at the distance $z=L$ downstream of the original object plane. This reconstruction can be achieved via the angular-spectrum diffraction integral, which is equivalent to the first Rayleigh-Sommerfeld diffraction integral for forward-propagating complex scalar waves, reducing to the Fresnel integral in the paraxial limit [1,2]. Thus we have the reconstruction formula:

$$E(x,y,z)\big|_{z=0} \approx \frac{1}{2\pi} \iint_{K_S} A_G(k_x, k_y) \exp(iL\sqrt{k^2 - k_x^2 - k_y^2}) \exp(-ixk_x - iyk_y) dk_x dk_y, \quad (1)$$

where the angular spectrum of the Gabor hologram is given by:

$$A_G(k_x, k_y) = \frac{1}{2\pi} \iint_{K_S} I_G(x, y, z = L) \exp(-ixk_x - iyk_y) dxdy. \quad (2)$$

Usually, both of these integrals are calculated by an FFT algorithm [1,2]. As result, the pixel size in the reconstructed images will be the same as in the hologram, i.e. the pixel size of the CCD camera. Nevertheless, the Abbe theoretical limit, $\delta$, for the resolution is described by the following relation:

$$\delta \geq \frac{0.5\lambda}{NA}, \quad (3)$$

where $\lambda$ is the wavelength and NA denotes the numerical aperture.

The pixel size of the CCD camera, used to register the Gabor hologram, may be larger than the theoretical resolution limit. In this case we may re-sample the image of the hologram, i.e. interpolate it onto a finer Cartesian grid, to increase the resolution of reconstructed images. In general, sharp details in the hologram are produced by large-scale features of the recorded object, and conversely, small-scale features produce extended smooth features in the hologram. The slowly varying information on the hologram can often be successfully interpolated between pixels, which corresponds to increasing the resolution of the small scale features in the reconstruction of the original object.

Accordingly, we propose to re-sample the digital inline holographic data from the CCD to increase the resolution of the reconstructed images. This procedure amounts to inverting an interpolated Fresnel diffraction image to recover the object. This technique has been observed to increase the resolution of the reconstructed images by 3-5 times, as shown in the experimental demonstration to which we now turn.

## 3. Experiment

To demonstrate our resampling method, we took inline holographic images of a microscope calibration slide (Motic, 10 μm spacing between the smallest lines) at a distance of 33 +/- 0.5 mm downstream of the object, using HeNe laser illumination (wavelength 632.8 nm). The pixel size of our CCD camera was 7.4 micrometers and without re-sampling we use Eqs (1) and (2) to obtain the reconstructed images shown in Fig.2a; in these reconstructions based on a non-re-sampled hologram, we cannot resolve the calibration mash with the period of 10 micrometers. However, if we use nearest-neighbour linear interpolation to re-sample the holographic images and increase the number of pixels by 3.5 times in both directions, then the calibration mesh is very well resolved in the associated numerical reconstruction based on Eqs (1) and (2) (see Fig.2b).

A schematic representation of the experimental setup is shown in Fig. 1. The beam from a HeNe laser (1 mW, random polarization) is attenuated by a neutral density filter (OD 2) before being sent through a spatial filter employing a 20 $\mu$m pinhole, with the exiting expanding beam re-collimated using a lens (f=250 mm) to produce an approximately planar wave.

This beam then illuminates the test object, and the resulting diffraction pattern is recorded in the near field by a monochrome CCD camera (Prosilica GE1650) with the CCD surface located at approximately 30 mm from the test object. The CCD array of the camera is 1600 (H) x 1200 (V) pixels, and each pixels is 7.4 $\mu$m square. To improve the signal to noise ratio of the recorded diffraction image, 100 background subtracted images are recorded and averaged, which allows the camera to approach its sensitivity of 12 bits.

The reconstructed object shown in Figs. 3a-c is the wing from a small Australian gnat (approximately 1.9mm x 0.7 mm in size) which was recorded at a distance of (31.1 +/- 0.5) mm downstream of the object. Figures 3a-b are reconstructed images and Fig. 3c is the original hologram. Due to the noticeable attenuation of the beam by all parts of the object, a high-dynamic-range diffraction image of this object was recorded. The image was assembled from four different sets of exposures, where the exposure times for each set differ by a factor of approximately 3.2. Each set consists of 100 background subtracted images which are subsequently averaged to produce a final image for the set. The final high dynamic range image is produced by starting with the longest exposed averaged image and replacing regions greater than 71% of the saturated value with values from the same region of the next-longest exposed image, weighted by the increased exposure time. This process is applied to the new composite image until information from the least exposed image is reached. The final dynamic range achieved in the final image is approximately 1:100000 or close to 17 bits. As Fig. 3b shows, the fine structure of the fly wing is better resolved on the re-sampled image.

**Conclusions**

We introduce a simple method of re-sampling inline-holographic data to increase the spatial resolution of the reconstructed images. The proposed technique was successfully validated using experimental data. This simple method may be applied to analyze the holographic data obtained using x-rays, electrons or visible light. It gives a very simple way to increase the spatial resolution of holographically recorded images of small objects. The method may be applied only in the Fresnel regime of diffraction. Re-sampling of the holographic data in the Fraunhofer regime does not increase resolution of the scheme presented here, but increases the viewed area of the object.

**References and links**

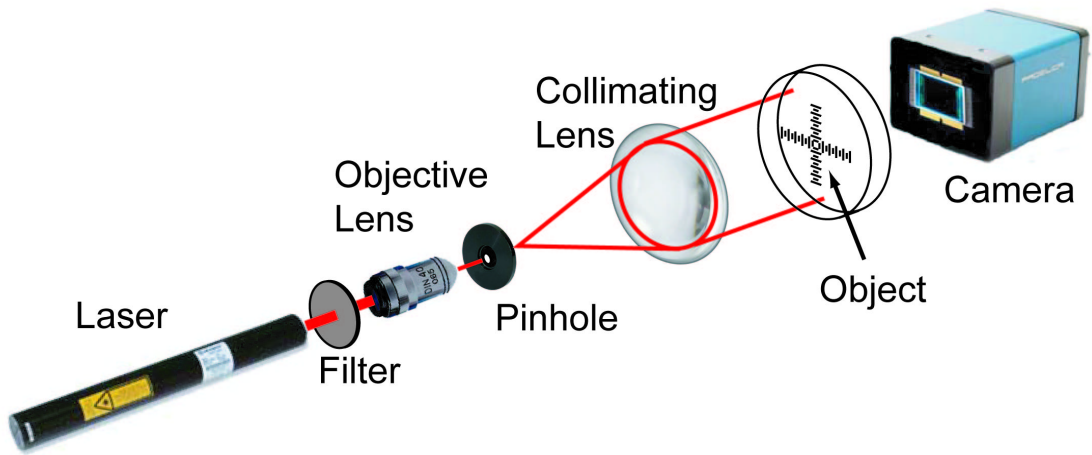

Fig. 1 Experimental setup for collection of Gabor inline holograms.

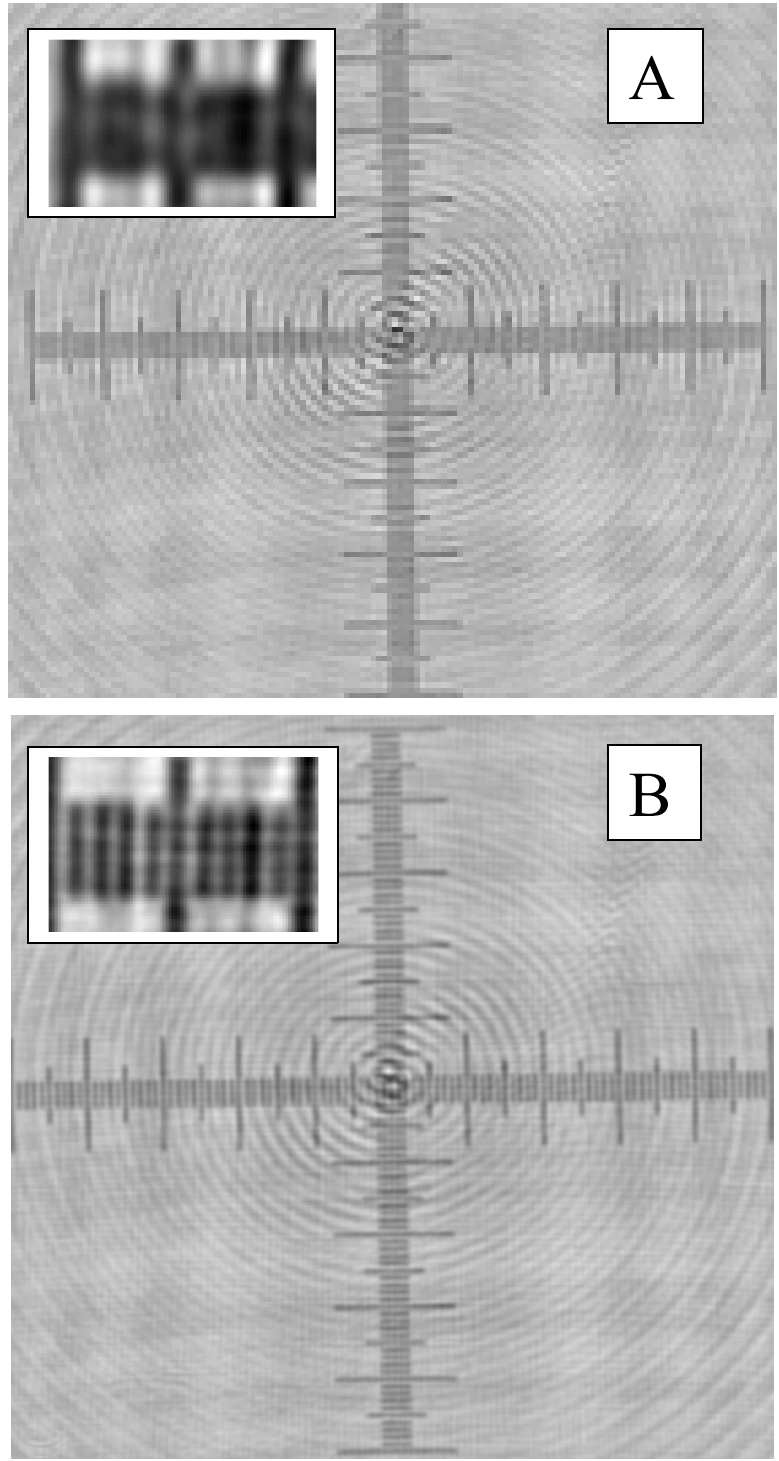

Fig. 2 A- reconstructed original image of the microscope calibration slide. The smallest unresolved bars are 10 µm apart, B-reconstructed re-sampled (3.5 x 3.5 times) image. The boxes show a magnified region of each image.

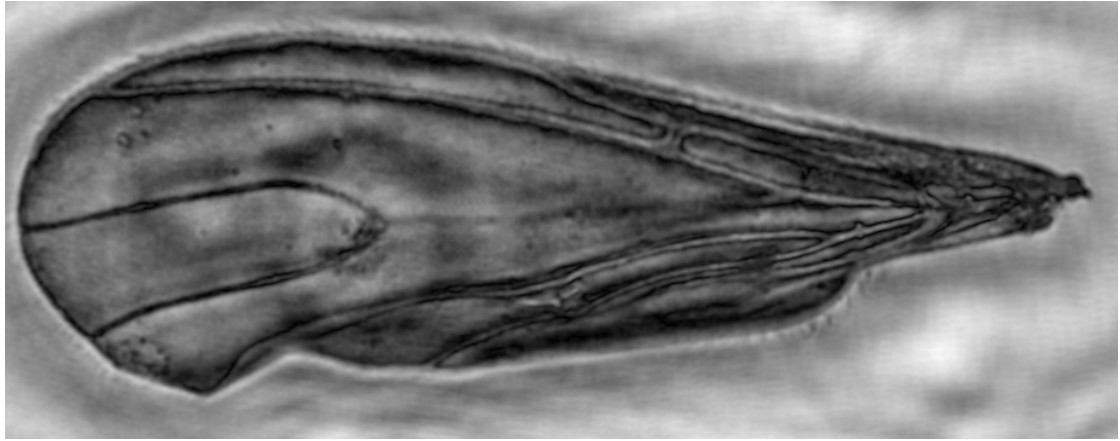

A)

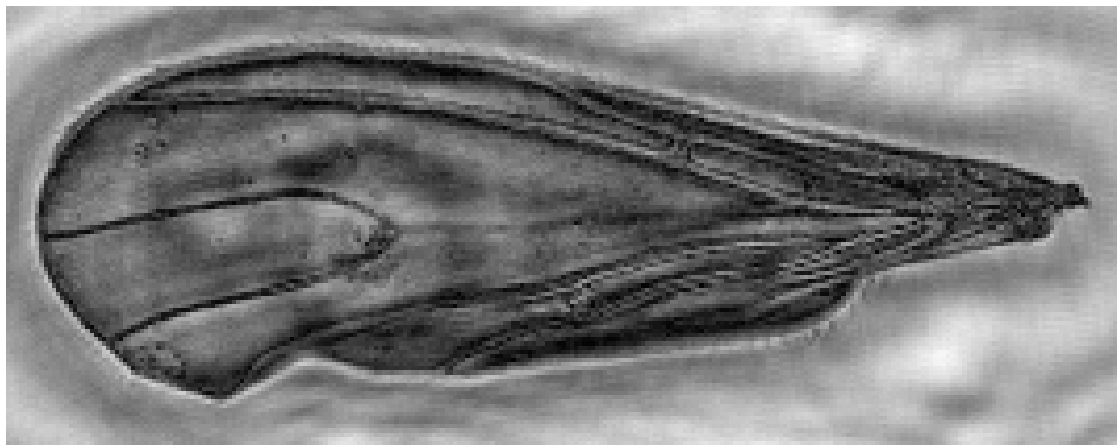

B)

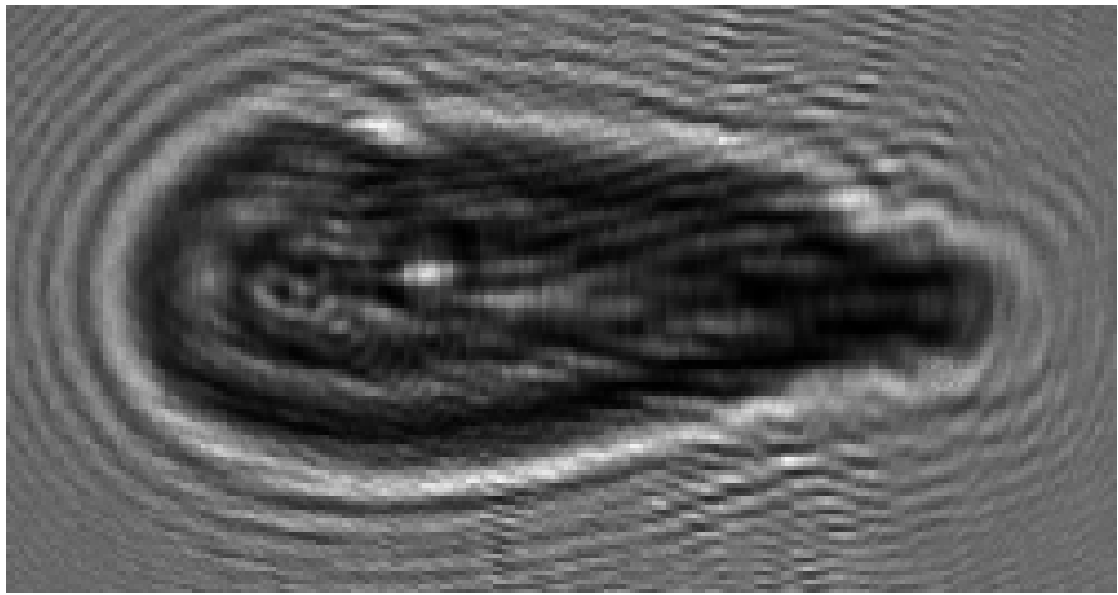

C)

Fig. 3 A) – reconstructed re-sampled (3.5 times ) image of fly wing (pixel size is 2.1 micrometers), B) – original reconstructed image (pixel size is 7.4 micrometers) and C) – their holographic image (part of hologram)